\documentclass[12pt]{iopart}

\usepackage{graphicx}
\usepackage{iopams}
\begin{document}

\title{Quasi-linear magnetoresistance and the violation of Kohler's rule in the quasi-one-dimensional Ta$_4$Pd$_3$Te$_{16}$ superconductor}

\author{Xiaofeng Xu$^1$, W. H. Jiao$^2$, N. Zhou$^1$, Y. Guo$^1$, Y. K. Li$^1$, Jianhui Dai$^1$, Z. Q. Lin$^3$, Y. J. Liu$^3$, Zengwei Zhu$^3$, Xin Lu$^4$, H. Q. Yuan$^4$, Guanghan Cao$^5$}


\address{$^{1}$Department of Physics and Hangzhou Key Laboratory of Quantum Matters, Hangzhou Normal University, Hangzhou 310036, China\\
$^{2}$School of Science, Zhejiang University of Science and Technology, Hangzhou 310023, China\\
$^{3}$Wuhan National High Magnetic Field Center, School of Physics, Huazhong University of Science and Technology, Wuhan, 430074, China\\
$^{4}$Center for Correlated Matter and Department of Physics, Zhejiang University, Hangzhou, 310058, China\\
$^{5}$Department of Physics, Zhejiang University, Hangzhou 310027, China\\}

\date{\today}

\ead{xiaofeng.xu@hznu.edu.cn}

\begin{abstract}
We report on the quasi-linear in field intrachain magnetoresistance in the normal state of a quasi-one-dimensional superconductor
Ta$_4$Pd$_3$Te$_{16}$ ($T_c$$\sim$4.6 K). Both the longitudinal and transverse in-chain magnetoresistance shows a power-law dependence,
$\Delta \rho$$\propto$B$^\alpha$, with the exponent $\alpha$ close to 1 over a wide temperature and field range. The magnetoresistance
shows no sign of saturation up to 50 tesla studied. The linear magnetoresistance observed in Ta$_4$Pd$_3$Te$_{16}$ is found to be overall
inconsistent with the interpretations based on the Dirac fermions in the quantum limit, charge conductivity fluctuations as well as quantum
electron-electron interference. Moreover, it is observed that the Kohler's rule, regardless of the field orientations, is violated in its
normal state. This result suggests the loss of charge carriers in the normal state of this chain-containing compound, due presumably to the
charge-density-wave fluctuations.
\end{abstract}

\maketitle

\section{Introduction}

Understanding the normal state properties of a superconductor is the key step to reveal its pairing mechanism\cite{Lee06}. However, the
normal state of some \textit{unconventional} superconductors, such as the high $T_c$ cuprates, may be sometimes more extraordinary and
controversial than their superconducting counterpart\cite{Hussey02}. In the pantheon of various superconducting materials,
quasi-one-dimensional (Q1D) superconductors hold a special place in the study of correlated electrons as they display many rich physical
phenomena exclusively for the reduced dimensionality, \textit{e.g.}, the field-induced spin density wave\cite{Kwak81}, a variety of angular
magnetoresistance oscillations\cite{Lebed,Osada,Danner94} and even spin-triplet pairing state\cite{Chaikin97}. Notably, the normal state of
some Q1D superconductors, including the organic Bechgaard salts (TMTSF)$_2$X(X=PF$_6$, ClO$_4$)\cite{Haddad04,zhang07} and purple bronze
Li$_{0.9}$Mo$_6$O$_{17}$\cite{Wakeham11}, is often regarded as the most promising candidate to realize the so-called Tomonaga-Luttinger
liquid (TLL) paradigm in the bulk materials.

In standard metals, the Lorentz force caused by an applied magnetic field changes the electron trajectory and gives rise to a positive
magnetoresistance (MR) which increases quadratically with the strength of the field\cite{Ashcroftbook}. However, there are few exceptions
where the MR may grow linearly with the field. For example, in the Dirac systems, the MR acquires the linear-in-field form once all Dirac
fermions are degenerate into the lowest Landau level, i.e., in the quantum limit regime\cite{Ong10,Wangkefeng11,BaFeAs11}. The linear MR
was also observed in some ferromagnets, including ferromagnet Fe$_{1-y}$Co$_y$Si crystals\cite{Manyala00} and geometrically constrained
thin films of iron, nickel and cobalt, due to quantum electron-electron ($e$-$e$) interference effects\cite{Gerber07}. Besides,
quasi-linear MR was also reported in non-magnetic silver chalcogenides\cite{Xu97}. Other mechanisms for the linear MR involve the
polycrystalline materials\cite{Kapitza1929} and inhomogeneous compounds with mixed components of the resistivity tensor\cite{Herring1960}.
On the other side, the MR at a certain temperature $\Delta \rho$ under a field $H$ obeys a general function known as the Kohler's rule:
$\Delta \rho$/$\rho_0$=$f$($H/\rho_0$), where $\rho_0$ is the zero-field resistivity\cite{Kohler1938,NieLuo02}. As a result, the plots of
$\Delta \rho/\rho_0$ as a function of $H/\rho_0$ at distinct temperatures will collapse onto a single curve. Interestingly, this rule,
although derived from the semiclassical Boltzmann theory, was found to be well obeyed in a large number of metals, including the metals
with two types of carriers, the pseudogap phase of the underdoped cuprates \cite{Greven14} as well as some other Q1D
metals\cite{Narduzzo07}. The violation of such a rule is generally believed to result from the loss of carriers with temperature or from
the fact that the anisotropic electron scattering rates do not have the same $T$ scaling on different sections of the Fermi surface (FS).

A new Q1D chain-containing compound Ta$_4$Pd$_3$Te$_{16}$ has recently been reported to be superconducting below $T_c$$\sim$4.6
K\cite{Jiao14}. In its crystal structure, one dimensional PdTe$_2$ \textit{conducting} chains are extended along the $b$-axis, sandwiched
by TaTe$_3$ chains and Ta$_2$Te$_4$ double chains. Soon after this finding, the low temperature thermal conductivity measurements revealed
the gap node in its order parameter\cite{SYLi14}, similar to the overdoped cuprate Tl$_2$Ba$_2$CuO$_{6+\delta}$. However, the scanning
tunnelling microscopy (STM) points out that its gap structure is more likely anisotropic without nodes\cite{HHWen14}. On the other hand,
both the $T_c$-pressure diagram and STM study suggest that the system is probably in the vicinity of an ordered state, presumably a
charge-density-wave (CDW) instability\cite{SYLi14,DLFeng15}. Recent density function calculations appear to rule out a magnetic instability
as the origin of this ordered state\cite{Singh14}. All these findings seem to suggest that our understanding of this Q1D superconductor is
far from complete. The normal state properties, which would provide valuable clues to its superconducting mechanism, have hardly been
studied thus far. In this context, we study the normal state transport properties of this new Q1D superconductor and uncover that its MR,
unlike most of standard metals, shows quasi-linear behaviors in a broad $T$ and field range. Additionally, the semiclassical Kohler's rule
is found to be modestly violated, in \textit{all} three field orientations studied here. The implication of our observation has been
discussed.

\section{Experiment}

Ta$_4$Pd$_3$Te$_{16}$ single crystals were synthesized by a solid state reaction in vacuum, following the same procedures described in Ref.
[24]. The as-grown crystals have a typical size of 2.5$\times$0.25$\times$0.1 mm$^3$, with the longest dimension parallel to the chain
direction ($b$-axis). X-ray diffraction (XRD) and dc magnetization measurements were performed to confirm the sample quality. The MR was
measured by a standard four-probe technique with the current flowing along the $b$-axis for different field orientations up to 9 tesla and
13 tesla in superconducting magnets, respectively, and up to 50 tesla in the pulsed magnetic field laboratory. In this study, at least four
crystals from the same growth batch were measured under different magnets and field orientations.

\section{Results And Discussion}

The schematic view of the crystal structure projected on the $ac$ plane is shown in Fig. 1. We define hereafter the $a^*$-axis as the
direction in the flat Ta-Pd-Te layer orthogonal to the chains and the $c^*$-axis perpendicular to the $a^*b$ plane. The representative plot
of the temperature dependence of the in-chain resistivity is given in Fig. 1, with a blow-up of its low-$T$ superconducting transition as
the inset. Clearly, the midpoint of a sharp superconducting transition occurs at $\sim$ 4.6 K. Due to the sample morphology, it is
impossible to measure directly the inter-chain resistivity along the other two orthogonal directions, $\rho_{c^*}$ and $\rho_{a^*}$.
Instead, we use the anisotropy in its upper critical field to evaluate its resistivity anisotropy. According to anisotropic Ginzburg-Landau
theory, the upper critical field $H_{c2}$ with a field applied along $i$ direction is $H_{c2}^i$=$\frac{\Phi_0}{2\pi \xi_j \xi_k}$, where
$\Phi_0$ is the fluxoid and $\xi_{j,k}$ ($\propto \upsilon_{j,k}$) is the coherence length in the directions orthogonal to the field. On
the other hand, the resistivity $\rho_i$ is inversely dependent on the square of the Fermi velocity,
$\rho_i$$\propto$$\frac{1}{\upsilon_i^2}$. These combined give the following relations:
$\frac{H_{c2}^i}{H_{c2}^j}$=$\frac{\xi_i}{\xi_j}$=$\frac{\sqrt \rho_j}{\sqrt \rho_i}$. Fig. 2 displays the temperature dependence of the
$b$-axis resistivity under several values of magnetic field applied along the three orthogonal directions, along with the cumulative phase
diagram of $H_{c2}$, using the criteria of the midpoint of the transition. The resultant $H_{c2}$ extrapolated to $T$=0 K is 5.4 T, 9.4 T
and 3.3 T, for $H\parallel a^*$, $H\parallel b$, $H\parallel c^*$, respectively, in agreement with previous
measurements\cite{SYLi14,Jiao15}. Therefore, the resistivity anisotropy $\rho_{a^*}$:$\rho_b$:$\rho_{c^*}$ is estimated to be 3:1:8. This
anisotropy is rather small compared with other Q1D materials, like (TMTSF)$_2$X(X=PF$_6$, ClO$_4$), Li$_{0.9}$Mo$_6$O$_{17}$ and
PrBa$_2$Cu$_4$O$_8$.

Fig. 3 shows a series of field sweeps at fixed temperatures in the normal state of one sample studied (hereafter labelled as $\sharp$1). In
$H\parallel$$a^*$ and $H\parallel b$ configurations, the field is swept up to 9 T, while for $H\parallel c^*$, the maximum field is 13 T.
It is evident that the MR is nonquadratic and can be best fitted to a power-law dependence, $\Delta \rho/\rho$$\propto$$B^\alpha$. While
the exponent $\alpha$ shows a slight increase with the increasing temperatures, it is close to 1 in the whole temperature range studied.
The magnitude of the MR is comparable when field is aligned along the $a^*$- and $c^*$-axes, and is the smallest with field pointing to the
chain direction. The linearity of the MR is optimal at temperatures near 30 K. It is worth noting that the quasi-linear MR is not likely
derived from the superconducting fluctuations to such high temperatures, as even in the under-doped high $T_c$ cuprates with strong
superconducting fluctuations, the fluctuations extend only to a temperature no more than 5 times of $T_c$\cite{Ong00,Rourke11}.

In order to see if this power-law like MR will ever saturate at a higher field, we study the transverse MR ($H \parallel c^*$) of a second
sample ($\sharp$2) in the pulsed magnetic field. As seen from Fig. 4, this quasi-linear MR shows no sign of saturation up to 50 T. In
general, the transverse MR of a metal will saturate in the high fields once $\omega_c \tau$$\gg$1 unless the material is perfectly
compensated or it has open orbit in the FS\cite{Ashcroftbook,Cava14}. In Ta$_4$Pd$_3$Te$_{16}$, according to band structure
calculations\cite{Singh14}, its FS contains a 2D hole cylinder '$\alpha$', two nested 1D sheets '$\beta$' and '$\gamma$', and a 3D sheet
'$\delta$'. Hence, the open orbit associated with the 1D Fermi sheets may be responsible for the nonsaturating MR observed here.

Fig. 5 collectively shows the exponent $\alpha$ in $B^\alpha$ dependence of the MR at different temperatures and field configurations from
four samples studied (the other two labelled as $\sharp$3 and $\sharp$4). There are total 48 points in this figure. The most striking
feature of this figure is that, most of the data points (especially below 50 K) resides in the range of $\frac{4}{5}$ to $\frac{4}{3}$,
although $\alpha$ increases slightly with temperature.

In recent years, linear MR has been widely observed in the Dirac materials, such as topological insulators\cite{Ong10}, 3D bulk materials
with Dirac states\cite{Wangkefeng11}, and iron-based superconductor BaFe$_2$As$_2$\cite{BaFeAs11}. The linear energy dispersion of Dirac
fermions leads to nonsaturated linear MR once the field exceeds a critical field $B^*$ such that all states occupy the lowest Landau level
in the quantum limit. Therefore, the transverse MR displays a crossover from the low-$B$ quadratic dependence to the high-$B$ linear MR at
the critical field $B^*$. The crossover field $B^*$ has the $T$-dependence: $B^*$=$\frac{1}{2e\hbar\upsilon_F^2}$($E_F+k_BT$)$^2$, where
$\upsilon_F$ and $E_F$ are Fermi velocity and Fermi energy respectively\cite{Wangkefeng11} In Ta$_4$Pd$_3$Te$_{16}$, however, its low-$B$
MR profile can not be fitted to the quadratic form in a reasonable field window. In addition, the above $T$-dependence of the critical
field is not observed either.

In silver chalcogenides, Ag$_{2+\delta}$Se etc., the resistance exhibits an unusually linear dependence on magnetic field without any signs
of saturation at fields as high as 60 T\cite{Manyala00,Husmann02}. The underlying physical origin of the linear MR in this non-magnetic
material remains controversial\cite{Abrikosov98,Littlewood03,WeiZhang11}. A plausible explanation is the conductivity fluctuations
associated with inhomogeneous distribution of silver ions\cite{Littlewood03}. In our Ta$_4$Pd$_3$Te$_{16}$ crystals, the sample is in
single phase and the quality is high, confirmed from both XRD and energy dispersive x-ray spectra, hence this mechanism is unlikely here.

As described earlier, the linear positive MR has also been observed in some ferromagnets, such as the cobalt-doped FeSi\cite{Manyala00}.
This linear MR was attributed to the quantum $e$-$e$ interference interaction. The effect of a magnetic field on the $e$-$e$ interaction
was derived three decades ago in the nonmagnetic cases\cite{PatrickLee82}. Under this circumstance, the magnetic field induces a spin gap
($\propto$$g\mu_B H$, where $g$ is the Lande factor and $\mu_B$ is the Bohr magneton) which suppresses the contribution of $e$-$e$
interaction to the conductivity and leads to a positive MR proportional to ln($g\mu_BH/k_BT$) in 2D and to $\sqrt{g\mu_BH}$ in
3D\cite{PatrickLee82}. In a material with ferromagnetic correlations, however, there exists an (exchange) gap in the absence of the
external field. The external field further increases the gap and induces a correction to the resistivity which is linearly proportional to
$H$ at any laboratory field. However, there is no observed experimental evidence to date in favor of such ferromagnetic correlations in
Ta$_4$Pd$_3$Te$_{16}$. Instead, both the pressure and STM study suggested that the material may be actually close to a CDW
instability\cite{SYLi14,DLFeng15}. The question therefore remains of how the electrons scatter off the CDW fluctuations and ultimately lead
to a linear MR seen in this study.

At last, let us examine the Kohler's rule in this Q1D material. Kohler's plot, as exemplified for sample $\sharp$1, is shown in Fig.
6(a)-(c) for three different field orientations respectively. Clearly, the Kohler's rule is violated, in particular below $\sim$50 K, in
all field directions studied here. Generally, the departure of the Kohler's scaling results from the loss of the carriers or the
anisotropic scattering $\tau (k)$ that does not have the same $T$-scaling on different portions of the FS. Prior to the report of Ref.
[22], Kohler's rule was widely believed to be violated in cuprate superconductors as a result from the two-lifetime scattering and
non-Fermi liquid excitations. In the disordered Q1D PrBa$_2$Cu$_4$O$_8$, Kohler's rule was seen to be violated whereas it was obeyed in the
pure, clean samples. This dichotomy was proposed to arise from the disorder-tuned dimensional crossover from 3D to pure 1D and the
corresponding spin-charge separation in the 1D TLL regime. Given the resistivity anisotropy quoted above, it is farfetched to assign
Ta$_4$Pd$_3$Te$_{16}$ into a TLL material. Instead, it is natural to deem that, as inspired by the pressure study\cite{SYLi14}, the
material is on the border of a CDW instability and the violation of Kohler's rule is the result from the gapping out of the electrons with
decreasing $T$ due to the density-wave formation\cite{DLFeng15}.

\section{Conclusion}

In conclusion, we have uncovered an anomalously quasi-linear MR, irrespective of the field directions, in the normal state of the Q1D
Ta$_4$Pd$_3$Te$_{16}$ superconductor. This quasi-linear MR shows nonsaturating behavior up to 50 T, the highest field in this study.
Moreover, the Kohler's rule was seen to be violated in all three field directions studied. In combination with the previous report of its
$T_c$-pressure diagram, it is tempting to link our observation to the proximity to a CDW instability. In this respect, it is interesting to
see how the Drude tail in the optical response evolves in the $T$ range studied here. The origin of the quasi-linear MR observed in this
study however invokes more theoretical investigations in the future.

\section{Acknowledgement}

The authors would like to thank C. Lester, C. M. J. Andrew, A. F. Bangura, Chao Cao, Zhixiang Shi, Shiyan Li for stimulating discussions.
This work is sponsored by the National Key Basic Research Program of China (Grant No. 2014CB648400), and by the NSFC (Grant No. 11474080,
11104051). X.X. would also like to acknowledge the financial support from the Distinguished Young Scientist Funds of Zhejiang Province (LR14A040001).\\

\pagebreak[5]

\begin{figure}
\begin{center}
\vspace{-.2cm}
\includegraphics[width=12cm,keepaspectratio=true]{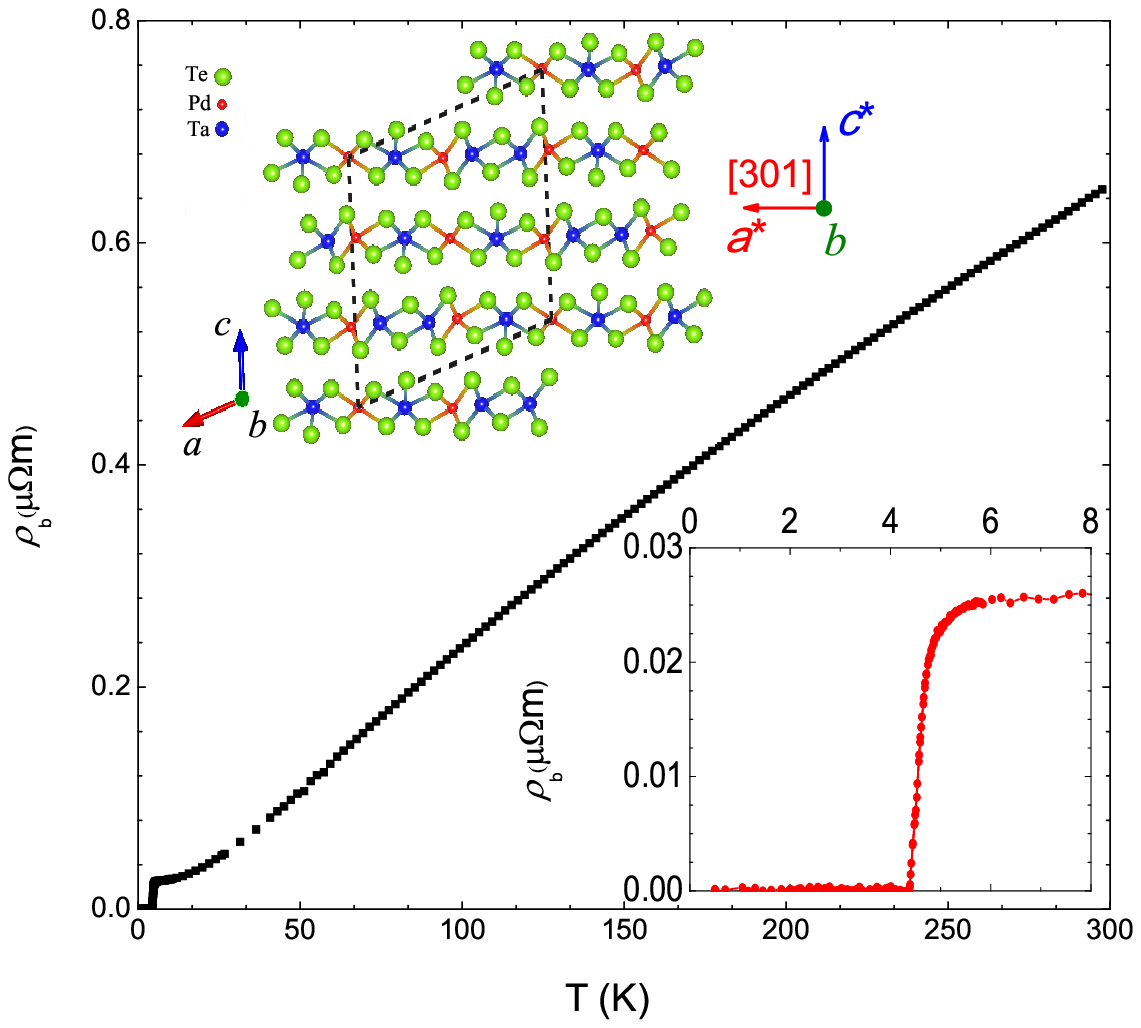}
\caption{(Color online) The representative zero-field resistivity as a function of temperature, with the bottom-right inset blowing up the
superconducting transition. Upper left inset: Schematic representation of the crystallographic structure of Ta$_4$Pd$_3$Te$_{16}$ as seen
from a perspective along the $b$-axis. Pd atoms (in red) are octahedrally coordinated by Te (in green), forming edge-sharing PdTe$_2$
chains along the $b$ axis. Te atoms display both prismy coordination and octahedral coordination around the Ta sites (in blue), forming
TaTe$_3$ chains and Ta$_2$Te$_4$ double chains. A new $a^*$-axis is defined in flat Ta-Pd-Te layers normal to the $b$-axis, as seen from
the inset and the $c^*$-axis is perpendicular to the $a^*$-axis. The unit cell is depicted by the thin dashed line. } \label{Fig1}
\end{center}
\end{figure}

\begin{figure}
\begin{center}
\vspace{-.2cm}
\includegraphics[width=18cm,keepaspectratio=true]{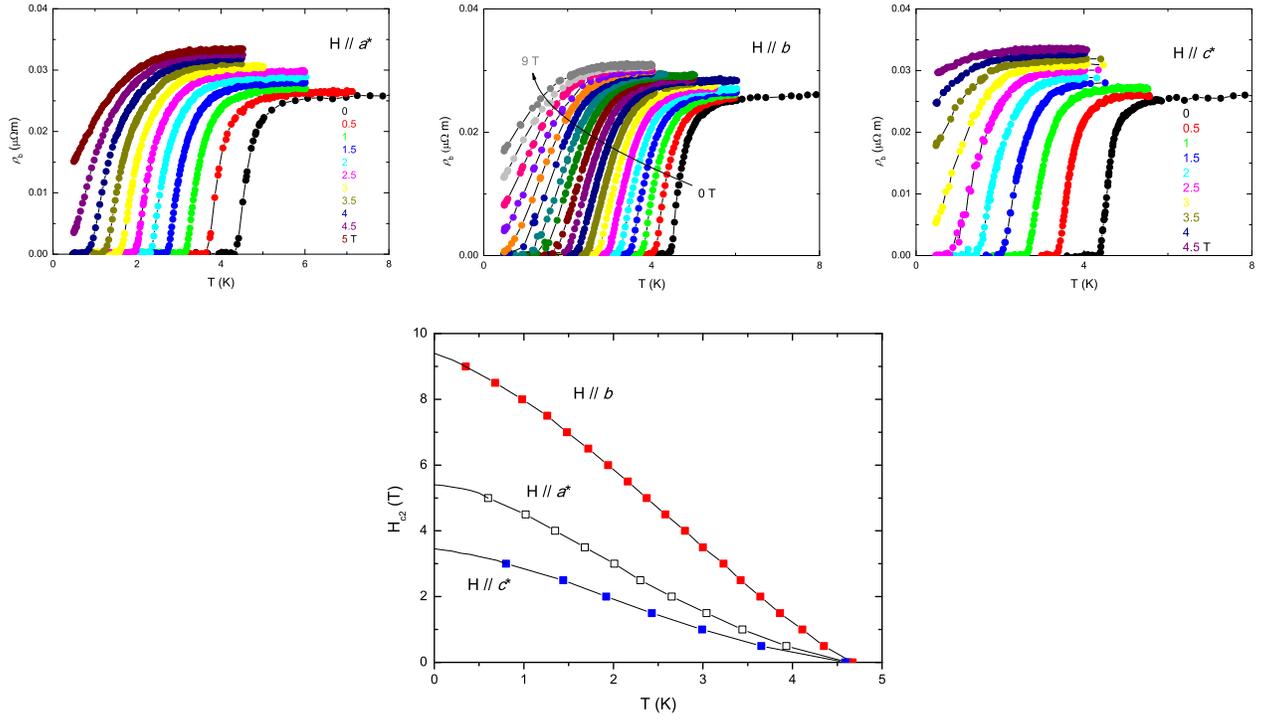}
\caption{(Color online) The top three panels show the $T$ sweeps of the $b$-axis resistivity of Ta$_4$Pd$_3$Te$_{16}$ under various fields
for $H$ aligned along the three orthogonal axes. Bottom panel: $H$-$T$ phase diagram using the midpoint of the transition for field aligned
along the $a^*$, $b$, and $c^*$ directions.} \label{Fig2}
\end{center}
\end{figure}

\begin{figure}
\begin{center}
\vspace{-.2cm}
\includegraphics[width=18cm,keepaspectratio=true]{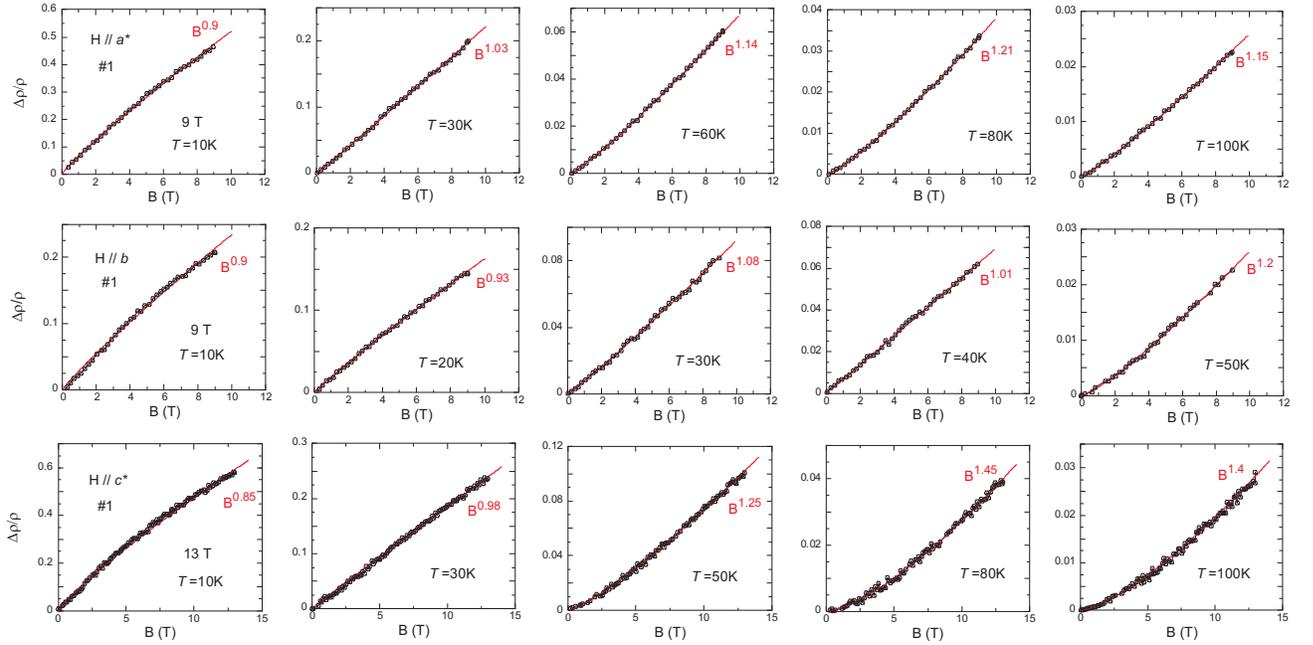}
\caption{(Color online) The field sweeps at several constant temperatures for $H\parallel a^*$, $H\parallel b$ and $H\parallel c^*$,
respectively, for sample $\sharp$1. For the former two field directions, the field is up to 9 T while for the latter, is up to 13 T. The
data are fitted to $B^\alpha$, with $\alpha$ given in the individual figures.} \label{Fig3}
\end{center}
\end{figure}

\begin{figure}
\begin{center}
\vspace{-.2cm}
\includegraphics[width=18cm,keepaspectratio=true]{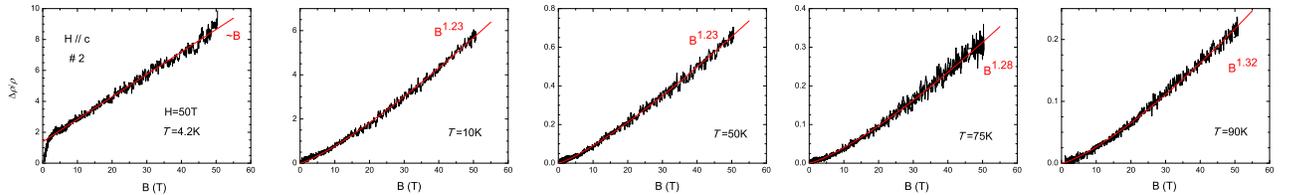}
\caption{(Color online) The $B$-sweeps at constant temperatures in the pulsed magnetic field along the $c^*$-axis up to 50 T. The data are
fitted to $B^\alpha$, with $\alpha$ indicated in each panel. The drop at low $B$ for $T$=4.2 K panel is due to superconductivity at this
temperature.} \label{Fig4}
\end{center}
\end{figure}

\begin{figure}
\begin{center}
\vspace{-.2cm}
\includegraphics[width=12cm,keepaspectratio=true]{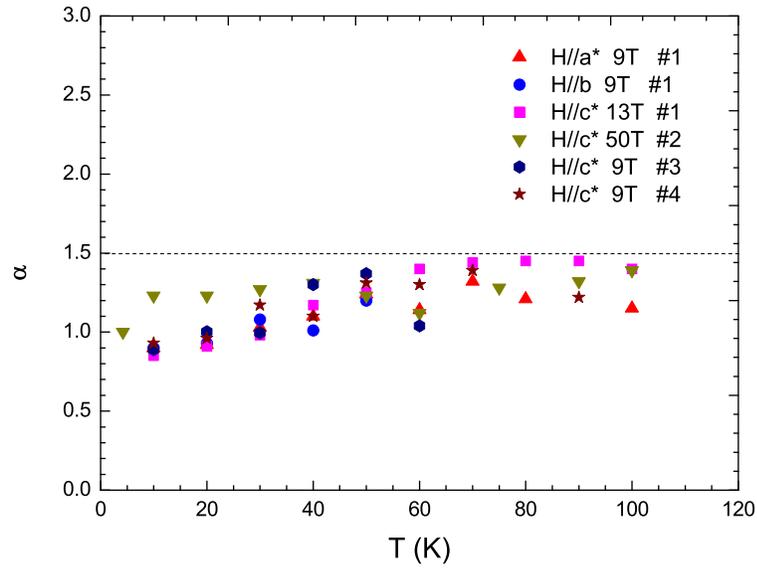}
\caption{(Color online) The exponent $\alpha$ in the power-law fitting $B^\alpha$, collected from 4 different samples under three different
magnets and field directions. Total 48 points are given in the plot.} \label{Fig5}
\end{center}
\end{figure}

\begin{figure}
\begin{center}
\vspace{-.2cm}
\includegraphics[width=9cm,keepaspectratio=true]{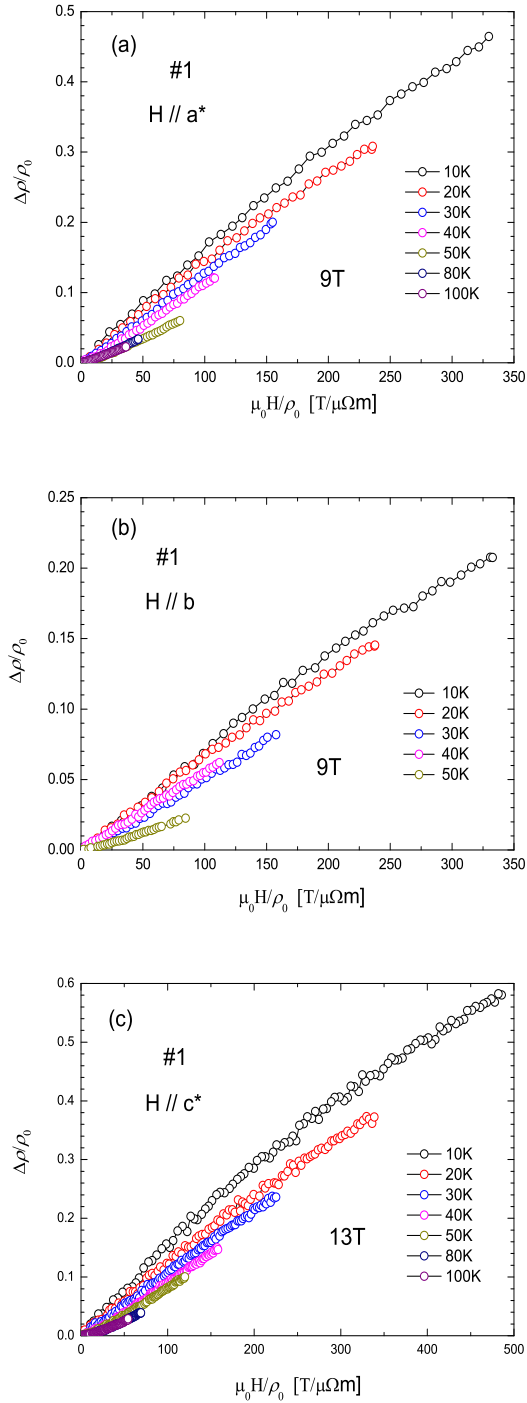}
\caption{(Color online) Panels (a)-(c) show the Kohler's plots for sample $\sharp$1 with $H\parallel a^*$, $H\parallel b$ and $H\parallel
c^*$, respectively. For $H\parallel a^*$, $H\parallel b$, the maximum field is 9 T and for $H\parallel c^*$, maximum field is 13 T. }
\label{Fig6}
\end{center}
\end{figure}

\end{document}